\documentstyle[12pt]{article}
\begin{document}
\thispagestyle{empty}
\null\vskip -1cm
\centerline{
\vbox{
\hbox{February 1996}\vskip -9pt
\hbox{hep-ph/9602423}\vskip -9pt
     }
\hfill 
\vbox{
\hbox{FERMILAB-PUB-96/027-T}\vskip -9pt
\hbox{UTEXAS-HEP-96-1}\vskip -9pt
\hbox{DOE-ER-40757-077} \vskip -9pt
\hbox{UICHEP-TH/96-7}\vskip -9pt
\hbox{UCD-96-3}\vskip -9pt
     }     } \vskip 1cm

\begin{center}
{\large \bf Color-Octet $J/\psi$ Production in the $\Upsilon$ Decay}

\vspace{0.1in}

Kingman Cheung\footnote{Internet address: {\tt cheung@utpapa.ph.utexas.edu}} 

{\it Center for Particle Physics, University of Texas at Austin, 
Austin, TX 78712} 

\vspace{0.1in}

Wai-Yee Keung\footnote{Internet address: {\tt keung@fnalv.fnal.gov}}

{\it Physics Department, University of Illinois at Chicago, IL 60607-7059\\
Fermilab, P.O. Box 500, Batavia, IL 60510}
\vspace {0.1in}

Tzu Chiang Yuan\footnote{Internet address: {\tt yuantc@ucdhep.ucdavis.edu}} 

{\it Davis Institute for High Energy Physics \\
University of California at Davis, Davis, CA 95616}

\end{center}

\begin{abstract}

The direct production rate of $\psi$ in the $\Upsilon$ decay is
shown to be dominated by the process  $ \Upsilon \to ggg^*$ followed by 
$g^* \to \psi$ via the color-octet mechanism proposed recently  
to explain the anomalous prompt charmonium production at the Tevatron.  
We show that this plausibly dominant process has a branching ratio 
compatible with the experimental data.  Further
experimental study in this channel is important to test the significance
of the color-octet component of $c\bar c$ pair inside the $\psi$ system.   

\end{abstract}
\newpage

\section{Introduction}
The most appealing explanation of the excessive production rates of 
prompt $\psi$, $\psi'$, and $\chi_{cJ}$ observed 
at the Fermilab Tevatron \cite{r:cdf1,r:cdf2} is given by the combination 
of the ideas of gluon fragmentation into quarkonium \cite{gfrag} 
and the color-octet mechanism \cite{r:braaten-fleming}, in which 
a gluon fragments into a color-octet $^3S_1$ $c\bar c$ pair 
which subsequently evolves nonperturbatively into the physical 
charmonium states by QCD dynamics.   
While the nonperturbative parameters associated with the color-octet mechanism 
must be extracted phenomenologically from the rates of prompt charmonium 
production, the prediction of the shape of the transverse momentum 
spectrum  agrees well 
with the data \cite{r:braaten-fleming,r:CGMP,r:cho-leibovich,r:bdfm}. 
A comprehensive review of these two theoretical developments 
and their implications at the Tevatron and LEP can be found in 
Ref.\cite{annrev}. 
If this mechanism is correct, it may give rise to many 
testable predictions for charmonium production in
$Z^0$ decay \cite{r:Z0}, 
low energy $e^+e^-$ annihilation \cite{r:ee}, 
photoproduction at fixed target and HERA experiments \cite{r:photo}, 
hadroproduction at fixed target experiments \cite{r:pN} 
and at LHC \cite{r:LHC}, and $B$-meson decays \cite{r:Bdecay}. 
Double prompt quarkonium production from the color-octet mechanism 
has also been studied at the Tevatron \cite{r:b-p-f}. 
In this paper, we show that the color-octet mechanism
can also provide the dominant contribution to $\psi$ production
in $\Upsilon$ decay. 

The available experimental data on charmonium production in 
$\Upsilon$ decay are  listed as follows,
\begin{equation}
   \hbox{Br} (\Upsilon \to \psi + X)
   \ \left\{
   \begin{array}{ll}
   = (1.1 \pm 0.4) \times 10^{-3} 
       & \mbox{{\small CLEO} \cite{r:CLEO},}    \\
   <  1.7 \times 10^{-3}
       & \mbox{Crystal Ball \cite{r:Crystal},}  \\ 
   <  0.68 \times 10^{-3}
       & \mbox{{\small ARGUS} \cite{r:ARGUS},}
   \end{array}
   \right.
\label{Bexp}
\end{equation}
in which the CLEO and the Crystal Ball results show a slight inconsistency.
But it is interesting to note that these experiments
can reach the branching fraction for $\Upsilon\to \psi + X$ at the
level of $10^{-4} - 10^{-3}$. 
Trottier \cite{r:trottier} studied the indirect  $\psi$ production
in $\Upsilon$ decay via the production of intermediate 
physical $\chi_c$ states, which decay radiatively into $\psi$.
However, this indirect production of $\psi$ contributes to a branching 
ratio less than $10^{-4}$ in $\Upsilon$ decay.  Since the experimental
branching fraction is already at the level of $10^{-4} - 10^{-3}$, 
we definitely need a new production mechanism to explain the data. 
We also hope that the slight discrepancy in the above experiments can be 
resolved in the near future. 

Conventional wisdom tells us that hadronically $\Upsilon$ decays predominantly 
through $b \bar b$ annihilation into three gluons. 
The rich  gluon content in the
final state makes it rather easy for the gluon to split into a $c \bar
c$ pair in the color-octet $^3S_1$ configuration. If $\psi$ can be formed
{}from this color-octet configuration at a significant level as predicted by 
Braaten and Fleming \cite{r:braaten-fleming} at the 
Tevatron \cite{r:cdf1,r:cdf2}, charmonium states should be 
abundantly produced in $\Upsilon$ decay.
A previous theoretical study \cite{r:Fritzsch} of the process  
$\Upsilon\to \psi +X$ was based on the color-evaporation model \cite{evapor}, 
with which the color-octet mechanism shares some common spirit, 
but the model fails to be systematic. 
Another qualitative estimate for $\Upsilon\to \psi +X$ can be found in 
Ref.\cite{r:Bigi}.

In Section 2, we will briefly review the 
description of the inclusive decay and production of quarkonium based on the 
nonrelativistic QCD (NRQCD) factorization formalism given by 
Bodwin, Braaten, and Lepage \cite{BBL}. 
In Section 3, we will discuss in detail several 
new color-octet processes relevant to $\psi$ production in $\Upsilon$ decay 
allowed by the general factorization formula. 
In Section 4, we compare the production rates of different 
processes and discuss the energy spectrum of $\psi$ in $\Upsilon$ decay.

\section{NRQCD Factorization Formalism}

The factorization formalism \cite{BBL} for the inclusive decay 
and production of heavy quarkonium 
allows us to probe the complete quarkonium Fock space in a systematic 
and consistent manner based upon NRQCD. It 
can be straightforwardly applied to the case of inclusive charmonium 
production from bottomonium decay \cite{r:trottier}. For the case of 
$\Upsilon \to \psi + X$, we have the following factorization formula,
\begin{equation}
\label{double_fact}
d \Gamma (\Upsilon \to \psi + X) = 
\sum_{m,n} d \widehat \Gamma_{mn} 
\langle \Upsilon \vert O_m \vert \Upsilon \rangle 
\langle  O_n^\psi  \rangle \; ,
\end{equation}
where $d \widehat \Gamma_{mn}$ are the short-distance factors for a $b \bar b$ 
pair in the state $m$ to decay into a $c \bar c$ pair in the state $n$ plus 
anything, where
$m,n$ denote collectively the color, total spin, and orbital 
angular-momentum of the heavy quark 
pairs.  $d \widehat \Gamma_{mn}$ can be calculated in perturbation theory as 
a series expansion in $\alpha_s(m_c)$ and/or $\alpha_s(m_b)$.
Contributions to $d \widehat \Gamma_{mn}$ that are sensitive to the quarkonium 
scales ($m_b v_b$ or $m_c v_c$ or smaller,  
where $v_b$ and $v_c$ are the relative velocities of the heavy quarks 
inside the bound states) and to $\Lambda_{QCD}$ 
can be absorbed into the NRQCD matrix elements 
$\langle \Upsilon \vert O_m \vert \Upsilon \rangle$ and  
$\langle  O_n^\psi  \rangle$.  
We use the notation $\langle  O_n^\psi  \rangle$ to denote the vacuum 
expectation value $\langle 0 \vert O_n^\psi \vert 0 \rangle$ of the operator
$O_n^\psi$.  
The nonperturbative factor 
$\langle \Upsilon \vert O_m \vert \Upsilon \rangle$ is proportional 
to the probability for the $b \bar b$ pair to be in the state $m$ inside the 
physical bound state $\Upsilon$, while 
$\langle  O_n^\psi  \rangle$ is proportional to 
the probability for a 
point-like $c \bar c$ pair in the state $n$ to form the bound state $\psi$. 
The relative importance of the various terms in the above double 
factorization formula (\ref{double_fact}) can be determined by the order 
of $v_{b}$ or $v_c$ in the NRQCD 
matrix elements and the order of $\alpha_s$ in the 
short-distance factors $d \widehat \Gamma_{mn}$. 

In the color-singlet model \cite{schuler}, 
the NRQCD matrix elements involved in the 
process $\Upsilon \to \psi + X$ are 
$\langle \Upsilon \vert O_1(^3S_1) \vert \Upsilon \rangle$ and 
$\langle  O_1^\psi (^3S_1)  \rangle$. 
According to the velocity scaling rules \cite{BBL}, they are scaled as
$m_b^3 v_b^3$ and $m_c^3 v_c^3$, respectively, and 
can be related to the quarkonium wavefunctions as follows:
\begin{eqnarray}
\label{wavefunction}
\langle \Upsilon \vert O_1(^3S_1) \vert \Upsilon \rangle
& \approx &  \frac{N_c}{2 \pi} \vert R_\Upsilon(0) \vert^2 \; \; ,\\
\langle  O_1^\psi (^3S_1)  \rangle  
& \approx & 3 \; \frac{N_c}{2 \pi} \vert R_\psi(0) \vert^2 \; \; ,
\end{eqnarray}
with $N_c$ denotes the number of colors. Therefore, these color-singlet 
matrix elements can be determined from the leptonic widths of the 
$\Upsilon$ and $\psi$. The short-distance factor in the 
color-singlet model for this direct process includes  
$b \bar b(^3S_1,\underline{1}) \to c \bar c(^3S_1,\underline{1}) gg$ and 
$b \bar b(^3S_1,\underline{1}) \to c \bar c(^3S_1,\underline{1}) gggg$. 
The two possible color configurations of the heavy quark pair are denoted 
by $\underline{1}$ for singlet and $\underline{8}$ for octet. 
The leading order Feynman diagrams for these processes are of 
order $\alpha_s^6$. 
Due to such a high order in the strong coupling constant, it is 
unlikely that these color-singlet processes can be the 
dominant production mechanism. 

The first example of the double factorization formula such as 
(\ref{double_fact})
is the indirect $\psi$ production in the 
decay $\Upsilon \to \chi_{cJ} + X$ with $\chi_{cJ} \to \psi\gamma$ 
considered by Trottier \cite{r:trottier}. 
The short-distance factor $\widehat \Gamma (b \bar b(^3S_1,\underline{1}) 
\to c \bar c(^3P_J,\underline{1})+ggg)$ is of order $\alpha_s^5$, which is 
enhanced by a factor of $1/\alpha_s$ compared with the direct 
color-singlet processes mentioned in the previous paragraph. 
However, the infrared divergence in the leading order calculation  
of the short-distance factor  indicates that the results 
are sensitive to the scale $m_c v_c$ or smaller. 
Therefore, in addition to the color-singlet matrix element 
$\langle  O_1^{\chi_{cJ}}(^3P_J)  \rangle$ 
(scales like $m_c^5 v_c^5$), one also needs to 
include the color-octet matrix element 
$\langle  O_8^{\chi_{cJ}}(^3S_1)  \rangle$ 
(scales like $m_c^3 v_c^5$) to absorb the infrared divergence.  
The short-distance process associated with the color-octet matrix element 
is $b \bar b(^3S_1,\underline{1}) \to c \bar c(^3S_1,\underline{8})+gg$, 
which is of order $\alpha_s^4$.  
In this case, the introduction of the 
color-octet matrix element is required by perturbative consistency, 
since the infrared divergence would otherwise 
spoil the one-term factorization formula. 

In the next section, we will consider the direct and indirect $\psi$ 
production in $\Upsilon$ decay with short-distance factors 
of order $\alpha_s^3$ and $\alpha_s^4$. 
These are possible only if higher Fock states of the $\psi$ or $\Upsilon$ 
are considered. We will also consider processes with short-distance 
factors of order $\alpha\alpha_s^3$, $\alpha^2\alpha_s$, 
and $\alpha^2\alpha_s^2$ that are  
suppressed by electromagnetic coupling but may or may not require higher 
Fock states of the quarkonia. 
In the following, we will first consider the produced 
$c \bar c$ pair in the color-octet 
$^1S_0$, $^3S_1$, or $^3P_J$ configuration, which subsequently evolves 
into physical $\psi$ described by the matrix elements 
$\langle  O^\psi_8 (^1S_0)  \rangle$, 
$\langle  O^\psi_8 (^3S_1)  \rangle$, or  
$\langle  O^\psi_8 (^3P_J)  \rangle$, respectively.
These color-octet matrix elements are suppressed by $v_c^4$ 
relative to the  color-singlet matrix element  
$\langle  O^\psi_1 (^3S_1)  \rangle$. 
The matrix element 
$\langle  O^\psi_8 (^3S_1)  \rangle$ has been extracted from the 
CDF data \cite{r:braaten-fleming,r:CGMP,r:cho-leibovich}, while 
two different combinations of the other color-octet matrix elements have   
been extracted from the CDF data \cite{r:cho-leibovich} and
{}from the photoproduction data by Amundson {\it et al} \cite{r:photo}.  

Though of much smaller effects, we also consider the contributions by the
higher Fock state of the color-octet $b \bar b$ pair 
inside the $\Upsilon$ associated with the matrix element 
$\langle \Upsilon \vert O_8 (^3S_1) \vert \Upsilon \rangle$, 
whose value has not yet been determined.  
An order of magnitude of this matrix element can, in principle, be estimated
by considering the ratio
\begin{equation}
\label{ratio1}
\frac{\langle \Upsilon |O_8(^3S_1) |\Upsilon \rangle }
     {\langle \Upsilon |O_1(^3S_1) |\Upsilon \rangle } \;
\frac{\langle  O_1^\psi(^3S_1)  \rangle }
     {\langle  O_8^\psi(^3S_1)  \rangle } \sim
\left( \frac{v_b^2}{v_c^2} \right)^2 \;, 
\end{equation}
which implies that its value should be highly suppressed. 
The ratio in (\ref{ratio1}) tells us that processes
associated with a color-octet $c \bar c$ pair inside the produced $\psi$ 
and a color-singlet $b \bar b$ pair inside the decaying $\Upsilon$ 
are much more important than
those with a color-octet $b \bar b$ pair inside the $\Upsilon$ and a 
color-singlet $c \bar c$ pair inside the $\psi$. 
Using the value $\langle  O_8^\psi(^3S_1) \rangle \approx 0.014$  
GeV$^{3}$ \cite{r:braaten-fleming,r:CGMP,r:cho-leibovich}, 
$\langle  O_1^\psi(^3S_1)  \rangle \approx 3
\langle \psi |O_1(^3S_1) |\psi \rangle \approx 0.73$ GeV$^3$ from the 
leptonic width of $\psi$, 
$\langle \Upsilon |O_1(^3S_1) |\Upsilon \rangle \approx 2.3$ GeV$^3$ from the 
leptonic width of $\Upsilon$, and $v_c^2\approx 0.3$ and $v_b^2\approx 0.08$,
we obtain $\langle \Upsilon |O_8(^3S_1) |\Upsilon \rangle \approx 
3 \times 10^{-3}$ GeV$^3$.  
However, such a large value for this matrix element would substantially
increase the hadronic width of $\Upsilon$, which would diminish the 
leptonic branching ratio to an unacceptable level.
Obviously, this matrix element enters into the hadronic width 
of the $\Upsilon$ via the short-distance process   
$b \bar b (^3S_1,\underline{8}) \to g^* \to q \bar q$.
In order not to spoil the experimental value for the leptonic branching ratio
and the total hadronic width of $\Upsilon$, it is 
necessary to put a bound on the value of the matrix element
$\langle \Upsilon |O_8(^3S_1) |\Upsilon \rangle$.  
The hadronic width of the $\Upsilon$ has the factored form
\begin{eqnarray}
\Gamma(\Upsilon \to {\rm light \; hadrons}) 
& = & \left( \hat \Gamma (b \bar b (^3S_1,\underline{1}) \to ggg) 
+  \sum_{q=u,d,s,c} \hat \Gamma (b \bar b (^3S_1,\underline{1}) 
\to q \bar q ) \right) 
\nonumber \\
&& 
\quad \times \langle \Upsilon \vert O_1(^3S_1) \vert \Upsilon \rangle 
\nonumber \\
&+& 
 \sum_{q=u,d,s,c} \hat \Gamma (b \bar b (^3S_1,\underline{8}) \to q \bar q )
\langle \Upsilon \vert O_8(^3S_1) \vert \Upsilon \rangle + \cdots \; ,
\end{eqnarray}
with  the following short-distance factors calculated to leading order in 
$\alpha$ and $\alpha_s$, 
\begin{equation}
\label{ggg}
\hat \Gamma(b \bar b(^3S_1,\underline{1}) \to ggg) = 
\frac{20\alpha_s^3}{243m_b^2} \; (\pi^2 - 9) \; \; ,
\end{equation}
\begin{equation}
\label{qqbar1}
\hat \Gamma(b \bar b (^3S_1,\underline{1}) \to f \bar f) 
=  \frac{2 \pi N_c Q_b^2 Q_f^2 \alpha^2}{3 m_b^2} \ ,
\end{equation}
and
\begin{equation}
\label{qqbar2}
\hat \Gamma(b \bar b (^3S_1,\underline{8}) \to q \bar q) 
= \frac{\pi \alpha_s^2}{3m_b^2} \ .
\end{equation}
In Eq.~(\ref{qqbar1}), $N_c$ is 1 and 3 for $f$ equals charged lepton $l$ and  
light quark $q$, respectively; 
$Q_f$ is the electric charge of the fermion $f$ in unit of the 
positron charge; and we have set $m_q$ and $m_l$ to zero in 
Eqs.~(\ref{qqbar1}) and (\ref{qqbar2}) for simplicity.  
Using the following expression for muonic branching ratio
\begin{equation}
{\rm BR}(\Upsilon \to \mu^+ \mu^-) = \frac{\Gamma(\Upsilon \to \mu^+ \mu^-)}
{\Gamma(\Upsilon \to {\rm light \; hadrons}) 
+\sum_\ell \Gamma(\Upsilon \to \ell^+ \ell^-) } \; ,
\end{equation}
together with the experimental value for BR$(\Upsilon \to \mu^+ \mu^-)=
0.0248 \pm 0.0007$ \cite{pdg}, we can obtain the following bound on 
$\langle \Upsilon |O_8(^3S_1) |\Upsilon \rangle$: 
\begin{equation}
\label{bound}
\langle \Upsilon |O_8(^3S_1) |\Upsilon \rangle \approx
\left(1.9 {  \begin{array}{c}  +5.1\\ -4.6 \end{array}  } \right) 
\times 10^{-4} \; {\rm GeV}^3  \ ,
\end{equation} 
where we have allowed a $2\sigma$ variation on BR$(\Upsilon \to \mu^+  \mu^-)$.
Alternatively, we can obtain another bound by 
using the total width of $\Upsilon$, but the result is not 
as good as the one given by Eq.(\ref{bound}).  Therefore, in the 
rest of the paper we will use the value 
$\langle \Upsilon |O_8(^3S_1) |\Upsilon \rangle 
\sim 5 \times 10^{-4}$ GeV$^3$.
With this value for $\langle \Upsilon |O_8(^3S_1) |\Upsilon \rangle$, 
we obtain the ratio
\begin{displaymath}
\frac{\hat \Gamma(b \bar b(^3S_1,\underline{8}) \to g^* \to q \bar q) 
\langle \Upsilon |O_8(^3S_1) |\Upsilon \rangle}
{\hat\Gamma(b \bar b(^3S_1,\underline{1}) \to ggg) 
\langle \Upsilon |O_1(^3S_1) |\Upsilon \rangle}
 \sim 0.017 \; ,
\end{displaymath}
which is now sufficiently small posing no threat to the experimental 
value of the leptonic width in the $\Upsilon$ decay.
The input parameters used in our later numerical calculations are summarized 
in Table~\ref{table1} for convenience. 
Since all our new calculations are at tree-level only, we, to be consistent, 
extract the color-singlet matrix elements from the leptonic widths of 
$\psi, \psi'$, and $\Upsilon$ using tree-level formulas.

\section{Color-Octet Processes}

We shall first consider two processes with a 
color-octet $c \bar c$ pair inside the produced $\psi$  and a color-singlet 
$b\bar b$ pair in the decaying $\Upsilon$. 

\subsection{$b \bar b (^3S_1,\underline{1}) \to \gamma^* \to c \bar c 
(^{2S+1}L_J,\underline{8}) +g$}

The leading order diagrams for the process 
$b \bar b (^3S_1,\underline{1}) \to c \bar c 
(^{2S+1}L_J,\underline{8}) +g$ is of order $\alpha^2 \alpha_s$, one of  
which  is shown in Fig.~\ref{fig1}(a).
This is similar to the process $Z \to c \bar c(^{2S+1}L_J,\underline{8})+ g$, 
which is negligible because the short-distance factor is suppressed by 
powers of $m_c^2/M_Z^2$ from the quark propagator \cite{r:Z0}. 
But in the present case it is only suppressed by powers of $m_c^2/m_b^2$.
We will restrict ourselves to the case of $L=0$ and 1 only. 
Although the contributions from higher values of $L$ can be included easily, 
they are further suppressed by powers of $v_c^2$. 
The inclusive production rate from these processes can be written as 
\begin{eqnarray}
\Gamma_{1a} (\Upsilon \to \psi + X) &=&
\langle \Upsilon \vert O_1(^3S_1) \vert \Upsilon \rangle \; \biggr[
\langle  O_8^\psi(^1S_0)  \rangle 
\widehat \Gamma (b \bar b (^3S_1,\underline{1}) \to c \bar c 
(^1S_0,\underline{8}) +g) \nonumber \\
&& \;\;\;\; + \sum_{J=0,1,2} \langle  O_8^\psi(^3P_J)  \rangle 
\widehat \Gamma (b \bar b (^3S_1,\underline{1}) \to c \bar c 
(^3P_J,\underline{8}) +g) \biggr] \; . 
\end{eqnarray}
The short-distance factors are calculated to leading order and are given by
\begin{eqnarray}
\widehat \Gamma(b \bar b(^3S_1,\underline{1}) \to c\bar c 
(^1S_0,\underline{8})+ g) &=&
 \frac{4 \pi^2 Q_c^2 Q_b^2 \alpha^2 \alpha_s}{3} \;
\frac{1}{m_b^4 m_c} \;
(1 - \xi ) \; , \\
\widehat \Gamma(b \bar b(^3S_1,\underline{1}) \to c\bar c
(^3P_0,\underline{8})+ g) &=&
 \frac{4 \pi^2 Q_c^2 Q_b^2 \alpha^2 \alpha_s}{9} \;
\frac{1}{m_b^4 m_c^3} \;
\frac{(1-3\xi)^2}{1 - \xi} \; , \\
\widehat \Gamma(b \bar b(^3S_1,\underline{1}) \to c\bar c
(^3P_1,\underline{8})+ g) &=&
 \frac{8 \pi^2 Q_c^2 Q_b^2 \alpha^2 \alpha_s}{9} \;
\frac{1}{m_b^4 m_c^3} \;
\frac{1+\xi}{1 - \xi} \; , \\
\widehat \Gamma(b \bar b(^3S_1,\underline{1}) \to c\bar c
(^3P_2,\underline{8})+ g) &=&
 \frac{8 \pi^2 Q_c^2 Q_b^2 \alpha^2 \alpha_s}{45} 
\frac{1}{m_b^4 m_c^3} \;
\frac{1+3\xi +6\xi^2}{1 - \xi} \; ,
\end{eqnarray}
where $\xi=M_\psi^2/M_\Upsilon^2\approx m_c^2/m_b^2$. 
We have also used the nonrelativistic approximation for the mass of the 
quarkonium: $M_\Upsilon \approx 2m_b$ and $M_\psi \approx 2m_c$. 
We note that 
$b \bar b (^3S_1,\underline{1}) \to \gamma^*  
\to c \bar c (^3S_1,\underline{8}) +g$ vanishes.
Using the heavy quark spin symmetry relation $\langle O_8^\psi(^3P_J) \rangle 
\approx (2J+1) \langle O_8^\psi(^3P_0) \rangle$ \cite{BBL}, 
the total width from these processes can be simplified as 
\begin{eqnarray}
\Gamma_{1a}(\Upsilon \to \psi + X) &=& \frac{4 \pi^2 Q_c^2 Q_b^2 \alpha^2 
\alpha_s}{3} \; 
\frac{\langle \Upsilon \vert O_1(^3S_1) \vert \Upsilon \rangle}{m_b^4 m_c} \;
\biggr\{ \langle  O_8^\psi(^1S_0)  \rangle (1-\xi) \nonumber \\
&& + 
 \frac{\langle  O_8^\psi(^3P_0)  \rangle }{3 m_c^2} \biggr[
\frac{(1-3\xi)^2}{1-\xi} +\frac{6(1+\xi)}{1-\xi} +
\frac{2(1+3\xi+6\xi^2)}{1-\xi} \biggr] \biggr\} \; ,
\end{eqnarray}
which can be normalized by the leptonic width of $\Upsilon$:
\begin{equation}
\label{lepton}
\Gamma(\Upsilon \to e^+ e^-) = \frac{2\pi Q_b^2 \alpha^2}{3}
\frac{\langle \Upsilon \vert O_1(^3S_1) \vert \Upsilon \rangle}{m_b^2} \; .
\end{equation}
Using $m_c = 1.5$ GeV, $m_b=4.9$ GeV, $\alpha_s(2m_b)=0.179$,
$\langle  O_8^\psi (^1S_0)  \rangle 
\approx \langle  O_8^\psi (^3P_0)  
\rangle /m_c^2 \approx 10^{-2}\, {\rm GeV}^3$ \cite{r:cho-leibovich},
and BR$(\Upsilon \to e^+ e^-) \approx 0.0252$ \cite{pdg}, 
the contribution from the above color-octet processes 
to the inclusive branching ratio BR$(\Upsilon \to \psi +X)$ is only 
$1.6 \times 10^{-5}$, which is 
almost two orders of magnitude below the CLEO data (\ref{Bexp}).  

\subsection{$b \bar b (^3S_1, \underline{1}) \to ggg^*
             \to c \bar c (^3S_1, \underline{8}) + g g$}

Fig.~\ref{fig1}(b) shows one of the six Feynman diagrams for the 
$b \bar b (^3S_1, \underline{1})$ 
pair annihilating into three gluons with one of the gluons converting into the 
$c \bar c (^3S_1, \underline{8})$ pair.  This process is of order 
$\alpha_s^4$ and its calculation is very much similar to 
the process $\Upsilon\to gg \gamma^* \to gg \ l\bar l$ \cite{r:leveille}.
Introducing the following scaling variables:
\begin{equation}
x_v = \frac{E_\psi}{m_b}, \qquad x_1 =\frac{E_{g_1}}{m_b}, 
\qquad x_2 =\frac{E_{g_2}}{m_b}, 
\end{equation}
such that $x_v + x_1 + x_2 =2$, where $E_i$ stands for the energy of the 
particle $i$. 
In the rest frame of $\Upsilon$, the
differential decay width is given by 
\begin{equation}
\frac{d \Gamma_{1b}}{d x_v d x_1} (\Upsilon \to \psi +X) = 
\frac{d \widehat \Gamma}{d x_v d x_1}
(b \bar b (^3S_1, \underline{1}) \to c \bar c (^3S_1, \underline{8}) + g g) 
\langle \Upsilon \vert O_1(^3S_1) \vert \Upsilon \rangle 
\langle  O_8^\psi (^3S_1)  \rangle \; \; ,
\end{equation}
with
\begin{eqnarray}
&&\frac{d \widehat \Gamma}{d x_v dx_1} 
 (b \bar b (^3S_1, \underline{1}) 
\to  c \bar c (^3S_1, \underline{8}) + g g) 
= 
\frac{5 \pi \alpha_s^4}{486 m_c^3 m_b^2}
 \;\frac{1}{ (x_v - 2 \xi)^2 x_1^2 (2 - x_v - x_1)^2}  \nonumber \\
&\times & \biggr [ 
2\xi^4 + 2\xi^3 (6 -4 x_v +2 x_1 -x_v x_1 -x_1^2 ) \nonumber \\
&+&  2\xi^2 \left( 11-16x_v +6x_v^2 - (8 -2x_v  -x_v^2) x_1 + (4 +x_v) x_1^2
  \right )  \nonumber \\
&+&  \xi \biggr( 4(1-x_v)(4-5x_v+2x_v^2)  - (32 -44 x_v 
   + 14 x_v^2) x_1 + (20  -18 x_v +x_v^2 ) x_1^2 \nonumber \\
&& \qquad - 2(2 - x_v) x_1^3 +x_1^4       \biggr) \nonumber \\
&+& 2 \biggr( 2 - 6x_v +7x_v^2 -4x_v^3 +x_v^4 - (6 -13 x_v +9 x_v^2  -
 2 x_v^3 )x_1 + (7 -9x_v  +3 x_v^2 )x_1^2 \nonumber \\
&& \qquad - 2(2 - x_v) x_1^3 +x_1^4 \biggr) \biggr ] \; , \label{junk}
\end{eqnarray}
where the ranges of integration for $x_v$ and $x_1$ are  
\begin{eqnarray}
\label{range1}
2\sqrt{\xi} \le & x_v & \le 1+\xi   \; , \\
\frac{1}{2}\left(2-x_v -\sqrt{x_v^2 -4\xi} \right ) \le & x_1 &  \le
\frac{1}{2}\left(2-x_v +\sqrt{x_v^2 -4\xi} \right ) \; .
\label{range2}
\end{eqnarray}
One can integrate over $x_1$ to obtain the energy distribution of $\psi$,  
\begin{eqnarray}
\label{edist1}
&& \frac{d \widehat \Gamma}{d x_v} 
(b \bar b (^3S_1, \underline{1})  \to  c \bar c (^3S_1, \underline{8}) + g g) 
= 
\frac{5 \pi \alpha_s^4}{486 m_c^3 m_b^2}
\frac{1}{ (2 - x_v)^3 (x_v - 2 \xi)^2} 
\nonumber\\
&\times & \biggr [ 
  4 (8+8\xi-14\xi^2-2\xi^3-12x_v-4\xi x_v+10\xi^2 x_v 
+5x_v^2-\xi x_v^2-\xi^2 x_v^2)  \nonumber \\
&& \times \ (1+\xi-x_v)  \log \left( 
\frac{2-x_v-\sqrt{x_v^2-4\xi}}{2-x_v+\sqrt{x_v^2-4\xi}} \right)
\  + \  (2-x_v)\sqrt{x_v^2-4\xi}
\nonumber \\
&& \times \ (16+28\xi+20\xi^2+4\xi^3-24x_v-36\xi x_v 
-12\xi^2x_v+14x_v^2+13\xi x_v^2-4 x_v^3)
\biggr ]   \; .
\end{eqnarray}
Numerically integrating $x_v$, we obtain the partial width, 
\begin{equation}
\label{width1}
\Gamma_{1b}(\Upsilon \to \psi +X) = 
\frac{5 \pi \alpha_s^4}{486 m_c^3 m_b^2}
\langle \Upsilon \vert O_1(^3S_1) \vert  \Upsilon \rangle 
\langle  O_8^\psi(^3S_1)  \rangle
 \; \times (0.571) \; .
\end{equation}
We note that the color-octet piece for the process 
$\Upsilon \to \chi_{cJ} + X$ considered by Trottier \cite{r:trottier} 
can be obtained from the above Eq.~(\ref{width1})  by simply replacing 
the matrix element 
$\langle  O_8^\psi(^3S_1)  \rangle$ with
$\langle  O_8^{\chi_{cJ}}(^3S_1)  \rangle$. 
However, the expressions for the energy distributions (\ref{junk}) and 
(\ref{edist1})
were not given explicitly in Ref.~\cite{r:trottier}. 
The prediction of the partial width is sensitive to the values  
of the two NRQCD matrix elements, the running coupling constant  
$\alpha_s$, and the heavy quark masses. 
For convenience we can normalize this partial width to the three-gluon width 
$\Gamma(\Upsilon \to ggg)$ given by Eq.~(\ref{ggg}) times 
the matrix element 
$\langle \Upsilon \vert O_1(^3S_1) \vert \Upsilon \rangle$:
\begin{equation}
\label{ratio}
\frac{\Gamma_{1b}(\Upsilon \to \psi +X)}{\Gamma(\Upsilon \to ggg)}
= \frac{\pi\alpha_s}{8}  
\frac{\langle  O^\psi_8(^3S_1)  \rangle}{m_c^3}  
\;\frac{0.571}{\pi^2 - 9}\; .
\end{equation}
It is insightful to take the scaling limit of $m_b \to \infty$ with 
$x_v=E_\psi/m_b$ held fixed in Eq.(\ref{edist1}). In this scaling limit, 
one can deduce 
\begin{equation}
\label{frag}
\frac{\Gamma_{1b}(\Upsilon \to \psi +X)}{\Gamma(\Upsilon \to ggg)} \approx 
\frac{\pi \alpha_s}{8} \frac{\langle  O^\psi_8(^3S_1)   
\rangle}{m_c^3} \;\; .
\end{equation}
The right-handed side of Eq.~(\ref{frag}) can be recognized 
to be three times the gluon fragmentation probability 
$P_{g \to \psi}$ in the color-octet mechanism obtained by Braaten and Fleming 
 \cite{r:braaten-fleming}. Thus, the above scaling limit corresponds  
to the fragmentation approximation. 
Comparing the above limit (\ref{frag}) with the exact result (\ref{ratio}), 
we see that fragmentation is not a good approximation. However,  
this limit suggests that 
the scale to evaluate the $\alpha_s$ in Eq.~(\ref{ratio}) 
should be $2 m_c$ instead of $2 m_b$. 
With $\alpha_s(2m_c)=0.253$, $\langle  O_8^\psi(^3S_1)   
\rangle = 0.014 \; {\rm GeV}^3$ \cite{r:braaten-fleming,r:cho-leibovich}, 
and assuming 
BR$(\Upsilon \to ggg)  \approx$ BR$(\Upsilon \to {\rm light \; hadrons}) 
= 0.92$ \cite{pdg}, we obtain 
\begin{eqnarray}
\frac{\Gamma_{1b}(\Upsilon \to \psi +X)}{\Gamma_{\rm total}(\Upsilon)} 
& = &
\frac{\Gamma_{1b}(\Upsilon \to \psi +X)}{\Gamma(\Upsilon \to ggg)}
\; \times \; {\rm BR}(\Upsilon \to ggg) 
\nonumber \\
& \approx & 2.5 \times 10^{-4} \; \; .
\end{eqnarray}
This  prediction is smaller than the CLEO data by merely a factor of 4, 
and is consistent with the bounds from Crystal Ball and ARGUS.

The color-octet process studied in this subsection applies to the case 
of $\psi'$ as well, simply by replacing the matrix element 
$\langle O_8^\psi (^3S_1) \rangle$ with 
$\langle O_8^{\psi'} (^3S_1) \rangle$, whose 
value has also been extracted from the CDF data to be 0.0042 
GeV$^3$ \cite{r:braaten-fleming,r:CGMP,r:cho-leibovich,r:bdfm}. 
With BR$(\psi' \to \psi + \gamma) \approx 57 \%$ \cite{pdg}, 
we obtain a branching 
ratio of $4.3 \times 10^{-5}$ for the inclusive $\psi$ production in the 
$\Upsilon$ decay that comes indirectly from $\psi'$. 

One can also consider the processes 
$\Upsilon \to \gamma g g^*$ followed by $g^* \to \psi \; (\psi')$ 
via the color-octet mechanism, and 
$\Upsilon \to gg \gamma^*$ followed by $\gamma^* \to \psi \; (\psi')$ 
in the color-singlet model.   
Up to overall normalization, the energy spectrum of the 
$\psi$ for these two processes are predicted to be same as 
in Eq.(\ref{edist1}). However, their partial widths  
are suppressed by factors of $8 \alpha/(15 \alpha_s) \sim 0.02$  and 
$32 \alpha^2 \langle O^\psi_1(^3S_1) \rangle
/(45 \alpha_s^2 \langle O^\psi_8(^3S_1) \rangle) \sim 0.06$, respectively,
compared with the width of Eq.(\ref{width1}). Thus they 
contribute a branching fraction about $2 \times 10^{-5}$ in the inclusive 
decay $\Upsilon \to \psi +X$. The indirect contribution from the $\psi'$ 
{}from these two processes is about $6\times 10^{-6}$.

\subsection{$b \bar b (^3S_1, \underline{8}) \to g^* \to 
c \bar c (^3P_J, \underline{1}) + g$}

We now turn to the case where the $b \bar b$ pair inside the $\Upsilon$ 
is in a color-octet $^3S_1$ state. The first process we consider 
is the production of $\chi_{cJ}$ from $\Upsilon$ decay. 
The leading order Feynman diagram is depicted in Fig.~\ref{fig1}(c). 
The factorization formula for the decay rate can be written as 
\begin{equation}
\Gamma_{1c} (\Upsilon \to \chi_{cJ} + X) = 
\widehat \Gamma (b \bar b (^3S_1, \underline{8}) \to  
c \bar c (^3P_J, \underline{1}) + g)
\langle \Upsilon \vert O_8(^3S_1) \vert \Upsilon \rangle 
\langle  O_1^{\chi_{cJ}}(^3P_J)  \rangle \; .
\end{equation}
We note that up to  coupling constants, color factors, 
and NRQCD matrix elements, these processes are similar to the ones
$b \bar b (^3S_1, \underline{1}) \to \gamma^* \to 
c \bar c (^3P_J, \underline{8}) + g$ considered in Section 3.1.   
The short-distance factors can be extracted easily from 
the previous calculations,   
\begin{eqnarray}
\widehat \Gamma
(b \bar b(^3S_1,\underline{8}) \to c \bar c(^3P_0,\underline{1})+g ) 
&=& \frac{\pi^2\alpha_s^3}{81} \;
 \frac{1}{m_b^4 m_c^3}\; 
 \frac{(1-3\xi)^2}{1-\xi} \; , \\
\widehat \Gamma
(b \bar b(^3S_1,\underline{8}) \to c \bar c(^3P_1,\underline{1})+g ) 
&=& \frac{2\pi^2\alpha_s^3}{81} \;
 \frac{1}{m_b^4 m_c^3}\; 
 \frac{1+\xi}{1-\xi} \; , \\
\widehat\Gamma
(b \bar b(^3S_1,\underline{8}) \to c \bar c(^3P_2,\underline{1})+g ) 
&=& \frac{2\pi^2 \alpha_s^3}{415} \;
 \frac{1}{m_b^4 m_c^3}\; 
 \frac{1+3\xi+6\xi^2}{1-\xi} \; .
\end{eqnarray}
The matrix element $\langle  O_1^{\chi_{cJ}}(^3P_J)  \rangle$ 
is related to the wavefunction according to \cite{BBL} 
\begin{equation}
\vert R'_{\chi_c} (0) \vert^2 \approx \frac{2 \pi}{9}
\frac{\langle  O_1^{\chi_{cJ}}(^3P_J)  \rangle}{2J+1} \; .
\end{equation}
Using  the value of the matrix element 
$\langle \Upsilon \vert O_8(^3S_1) \vert \Upsilon \rangle \approx 
5 \times 10^{-4}$ GeV$^3$ estimated in the last Section,  
$\alpha_s(2 m_b)=0.179$, 
$|R'_{\chi_c}(0)|^2=0.075 \hbox{GeV}^5$
{}from the potential model calculation \cite{quigg}, 
and the branching ratios 
BR$(\chi_{c1} \to \psi+\gamma)=0.273$ and BR$(\chi_{c2} \to \psi+\gamma) 
= 0.135$ ($\chi_{c0}$ has a negligible branching ratio 
into $\psi$) \cite{pdg}, 
we obtain the width $\Gamma_{1c}(\Upsilon \to \psi + X) \approx 0.05$ eV
and thus a branching ratio of  $9\times 10^{-7}$. 
Therefore, these indirect contributions  are negligible when
compared with the indirect mechanism considered earlier by 
Trottier \cite{r:trottier}.  

\subsection{$b \bar b (^3S_1,\underline{8}) \to g^* \to 
c \bar c (^3S_1,\underline{1}) + gg$}

One of the six leading Feynman diagrams 
for the short-distance process $b \bar b (^3S_1,\underline{8}) \to g^* \to 
c \bar c (^3S_1,\underline{1}) + gg$
is shown in Fig.~\ref{fig1}(d). 
The corresponding differential width can be written as 
the following factored form
\begin{equation}
\frac{d\Gamma_{1d}}{d x_v dx_1} (\Upsilon \to \psi + X)
= 
\frac{d\widehat\Gamma}{d x_v dx_1}
(b \bar b (^3S_1,\underline{8}) \to c \bar c (^3S_1,\underline{1})  gg )
\langle \Upsilon \vert O_8(^3S_1) \vert \Upsilon \rangle 
\langle  O_1^\psi (^3S_1)  \rangle 
\end{equation}
with the short-distance factor given by 
\begin{eqnarray}
\frac{d\widehat\Gamma}{d x_v dx_1} 
(b \bar b (^3S_1,\underline{8}) && \hskip -.5cm
  \to  c \bar c (^3S_1,\underline{1})  gg )
= \frac{5 \pi \alpha_s^4}{486 m_b^4 m_c} \,
\frac{1}{(2 - x_v)^2 (1 - \xi - x_1)^2 (1 +\xi - x_v - x_1)^2} 
\nonumber \\
&& \times \biggr \{ \xi^4 +2\xi^3 (8 -5x_v +x_v^2 -2x_1 +x_v x_1 +x_1^2)
\nonumber \\
&& + \xi^2 \biggr( 28 -46x_v +21 x_v^2 -4x_v^3 
		- (28 -26 x_v +6x_v^2) x_1
                   + (14  -6x_v) x_1^2 \biggr ) \nonumber \\
&& +2\xi \biggr( 
6-17x_v +16 x_v^2 -6x_v^3 +x_v^4 - (12 -22 x_v +12 x_v^2 - 2x_v^3) x_1
 \nonumber \\
&& + (10  -  12 x_v  +3 x_v^2 )x_1^2 - 2(2 -x_v) x_1^3 +x_1^4 
\biggr)\nonumber\\
&& -(1-x_1)(1-x_v -x_1)(1-x_v +2x_1 -x_v x_1 -x_1^2 ) 
			\biggr \} \;\; ,
\end{eqnarray}
where the allowable ranges of $x_v$ and $x_1$ are given in Eqs.~(\ref{range1})
and (\ref{range2}).
Integrating over $x_1$, we obtain the energy distribution for $\psi$ 
\begin{eqnarray}
\frac{d \widehat \Gamma}{d x_v } 
(b \bar b (^3S_1,\underline{8}) && \hskip -.5cm \to 
 c \bar c (^3S_1,\underline{1}) + gg )
= \frac{5 \pi \alpha_s^4}{486}\;
\frac{1}{m_b^4 m_c} \; 
\frac{1}{(2-x_v)^2(x_v-2 \xi)^3} \nonumber \\
&& \times \biggr \{
    \left( 4 +20\xi  + 28\xi^2 + 16\xi^3 - 12x_v - 
      36\xi x_v - 24\xi^2 x_v 
      13x_v^2 + 14\xi x_v^2 - 4x_v^3 \right)  \nonumber \\
&& \quad\times \left( x_v - 2\xi  \right)  \sqrt{x_v^2 - 4\xi}
   -4 \log \left( 
    \frac{2\xi  - x_v + \sqrt{x_v^2 -4\,\xi}}
	{2\xi  - x_v - \sqrt{x_v^2 -4\,\xi}}  
		\right)                        \nonumber \\
&& \quad\quad\times \left( 2\xi  + 16\xi^2 + 6\xi^3 - 16\xi^4 - 
      8\xi^5 - 12\xi x_v - 20\xi^2 x_v + 
      24\xi^3 x_v + x_v^2 \right. \nonumber \\ 
&& \quad\quad\quad +20\xi^4 x_v  + 12\xi x_v^2 - 8\xi^2 x_v^2  
   - 17 \xi^3 x_v^2 - x_v^3 - 
 \left.     \xi x_v^3 + 5 \xi ^2 x_v^3 \right)   \biggr \} \, . 
\label{edist2}
\end{eqnarray}
The partial width is obtained numerically: 
\begin{equation}
\label{width2}
\Gamma_{1d}(\Upsilon \to \psi +X) = 
\frac{5 \pi \alpha_s^4}{486}\;
\frac{\langle \Upsilon \vert O_8 (^3S_1) \vert \Upsilon \rangle}{m_b^4} \; 
\frac{\langle  O^\psi_1(^3S_1)  \rangle}{m_c} 
\; \times (1.230) \;\; .
\end{equation}
With the previous inputs, $\alpha_s(2m_b)=0.179$, and 
$\langle O_1^\psi(^3S_1) \rangle \approx 0.73$ GeV$^3$ obtained from the 
leptonic width of $\psi$,  we obtain a width of 0.02 eV only, which gives 
a branching ratio of $ 3 \times 10^{-7}$. 

\subsection{Other Color-Octet Processes}

When $\Upsilon$ annihilates into a $q\bar q$ pair via the $s$-channel photon 
$\gamma^\ast$, a bremsstrahlung virtual gluon emitted from the light 
quark line can become an octet $c\bar c(^3S_1,\underline{8})$, 
which then turns into $\psi$. The factored form of the rate for this process is
\begin{equation}
\Gamma_{1e}(\Upsilon\to \psi +X) = 
\hat \Gamma (b \bar b (^3S_1, \underline{1}) 
\to c \bar c (^3S_1, \underline{8}) + q\bar q) 
\langle \Upsilon \vert O_1(^3S_1) \vert \Upsilon \rangle \,
\langle O_8^\psi (^3S_1) \rangle \; .
\end{equation}
This process is of order $\alpha^2\alpha_s^2$, which is similar to a 
potentially dominant color-octet process in the prompt $\psi$ production in 
$Z^0$ decay \cite{r:Z0}. 
We can easily translate the formula from Ref.\cite{r:Z0} to the present case. 
The partial width is given by  
\begin{eqnarray}
\label{e:dl}
{\Gamma_{1e} (\Upsilon\to \gamma^* \to q \bar q \psi)
\over 
\Gamma(\Upsilon \to \gamma^* \to q\bar q)}    
&=&
\frac{\alpha_s^2(2m_c)}{36} \frac{\langle O_8^{\psi} (^3S_1) \rangle }{m_c^3} 
\bigg\{5(1-\xi^2)-2\xi\log\xi+\bigg[
 2\,\hbox{Li}_2\left({\xi\over 1+\xi} \right)
\nonumber \\
& - & 2\,\hbox{Li}_2\left({ 1 \over 1+\xi}\right)
-2\log(1+\xi)\log\xi + 3\log\xi+\log^2\xi\bigg](1+\xi)^2\bigg\} \, .
\nonumber \\
&& \;
\end{eqnarray}
Here Li$_2(x)=-\int_0^x{dt\over t}\log(1-t)$ is the Spence function and 
$\xi = (m_c/m_b)^2$. $\Gamma(\Upsilon \to q \bar q)$ is given by 
Eq.(\ref{qqbar1}) times the matrix element 
$\langle \Upsilon \vert O_1(^3S_1) \vert \Upsilon \rangle$. 
Using $m_b = 4.9$ GeV and $m_c = 1.5$ GeV,  $\xi \approx 0.0937$ and 
the curly bracket in (\ref{e:dl}) is about 1.1, as compared to 
the much larger value of 27.9 in the corresponding case in $Z^0$ decay
 \cite{r:Z0} where $\xi = m_\psi^2/M_Z^2 \approx 1.1 \times 10^{-3}$.  
Therefore,  the double logarithmic terms in  
Eq.(\ref{e:dl}) do not provide sufficient enhancement in our present case. 
Numerically, the ratio in (\ref{e:dl}) is only $8.3 \times 10^{-6}$, and so
this process can be safely ignored.

In addition to the above processes, 
one can also consider a color-octet $^3S_1$ 
$b \bar b$ pair annihilating into a light quark pair via a $s$-channel gluon, 
followed by an off-shell photon bremsstrahlung off either the $b$ quark line 
or the 
light quark line and eventually turns into the $\psi$.  This process 
also involves the matrix element 
$\langle \Upsilon \vert O_8(^3S_1) \vert \Upsilon \rangle$ 
and, therefore,  will be suppressed. 
One can also consider both the $b \bar b$ and $c \bar c$ pairs inside the 
quarkonia are in the color-octet configuration. Processes associated with 
such configuration are further suppressed by powers of $v_c^2$ and $v_b^2$ 
compared to those we have studied in this paper. We ignore them in this
work.

\section{Discussions and Conclusions}

The energy distribution of charmonium in the $\Upsilon$ decay can provide an
interesting test for the NRQCD factorization formalism 
discussed in Section 2. 
The energy distributions for the processes of Figs.~\ref{fig1}(a) and 
\ref{fig1}(c) are just a delta function with the peak at one half of the 
$\Upsilon$ mass, while
the energy distributions  
for the processes of Figs.~\ref{fig1}(b) and \ref{fig1}(d) are given by 
Eqs.~(\ref{edist1}) and (\ref{edist2}), respectively, 
and are shown in Fig.~\ref{fig2}. 
The solid curve in Fig.~\ref{fig2} 
is the energy spectrum of $\psi$ for the dominant 
process $b \bar b (^3S_1, \underline{1}) \to g^*gg \to   
          c \bar c (^3S_1, \underline{8})  gg$; 
it is monotonically increasing as the $\psi$ energy increases and eventually 
cut off by the kinematic limit.  
In the fragmentation approximation  
($m_b \to \infty$ with $E_\psi/m_b$ held fixed), 
the energy distribution of $\psi$ 
for this subprocess is the same as that of the fragmenting gluon, 
up to an overall constant.  The result using the fragmentation approximation 
is shown as the dashed curve in Fig.~\ref{fig2} for comparison. 
The fragmentation approximation has overestimated the exact result  
for all energies of the $\psi$. 
On the other hand, CLEO \cite{r:CLEO} obtained a relatively flat momentum 
spectrum.   However, the rising spectrum shown by the solid curve in 
Fig.~\ref{fig2} can be softened by various mechanisms such as 
final state interactions of the soft gluons \cite{r:photon} and by
relativistic corrections of the bound state \cite{r:yee}, just to name a few 
possibilities.   
This situation is very similar to the photon spectrum in the decay 
$\Upsilon \to \gamma gg$ \cite{r:photon,r:yee,r:crystal}. 
The dashed-dotted curve in Fig.~\ref{fig2} is the energy distribution for 
the suppressed channel   
$b \bar b (^3S_1, \underline{8}) \to g^*  \to  
	      c \bar c (^3S_1, \underline{1})  gg$ and it 
has a much flatter shape.  

Among the several processes that we have considered in this paper, 
those two with the color-octet $b \bar b$ pair inside the decaying $\Upsilon$
are more than 
two orders of magnitude below the CLEO data and therefore negligible, 
while the other with the color-octet $c \bar c$ pair 
inside the produced $\psi$ are relatively more important.  
The largest contribution is being the process shown in Fig.~\ref{fig1}(b),
which has a branching ratio about $2.5 \times 10^{-4}$. 
The next largest contribution is the indirect process considered by 
Trottier \cite{r:trottier}.   From Eq.(21) of Ref.~\cite{r:trottier} 
and using the heavy quark spin symmetry relation 
$\langle O_8^{\chi_{cJ}} (^3S_1)  \rangle \approx 
(2J+1) \langle O_8^{\chi_{c0}} (^3S_1)  \rangle$ \cite{BBL}
and the value of the matrix
element $\langle O_8^{\chi_{c0}} (^3S_1)  \rangle = 0.0076$ GeV$^3$
obtained by fitting the Tevatron data \cite{r:CGMP,annrev}, we obtain 
$\sum_J$BR$_{\rm Trottier}(\Upsilon \to \chi_{cJ} + X; \,
\chi_{cJ} \to \psi \gamma) \approx 5.7 \times 10^{-5}$.  
Processes that are comparable to this one are 
(1) the process shown in Fig.~\ref{fig1}(a) which has a branching ratio 
of $1.6 \times 10^{-5}$, 
(2) the processes $\Upsilon \to gg \gamma^* \to \psi +X$ and 
$\Upsilon \to \gamma g g^* \to \psi + X$ via the color-octet mechanism 
which have a combined branching ratio about $2 \times 10^{-5}$, 
and finally, (3) the indirect contribution from $\psi'$ 
having a branching ratio about $5 \times 10^{-5}$. 
Therefore, adding up the contributions from all these processes,  
we obtain a branching ratio BR$(\Upsilon \to \psi + X)  \approx 
4 \times 10^{-4}$, which is within $2\sigma$ of the CLEO data.  
Given these theoretical results, it would be very interesting to have  
more precise measurements of the inclusive rates and energy spectra of
charmonium from the $\Upsilon$ decay. 
This would provide a crucial test of the 
NRQCD factorization formalism applied simultaneously 
to both the bottomonium and charmonium system. 

This work was supported in part by the United States Department of 
Energy under Grant Numbers DE-FG02-84ER40173, DE-FG03-93ER40757,  
and DE-FG03-91ER40674.


\eject
\begin{table}[h]

\caption[]{\label{table1} \small Input parameters used in our calculations.}
\centering
\begin{tabular}{|c|c|}
\hline
\hline
NRQCD matrix elements &  Value \\
\hline
$\langle  O_1^\psi(^3S_1)  \rangle \approx 3
\langle \psi |O_1(^3S_1) |\psi \rangle$ &  0.73 GeV$^3$ \\
$\langle  O_1^{\psi'}(^3S_1)  \rangle \approx 3
\langle \psi' |O_1(^3S_1) |\psi' \rangle$ &  0.42 GeV$^3$ \\ 
$|R'_{\chi_c}(0)|^2 \approx 
\frac{2\pi}{9} \langle O_1^{\chi_{c0}}(^3P_0) \rangle$  & 0.075 GeV$^5$ 
\cite{quigg}\\
$\langle \Upsilon |O_1(^3S_1) |\Upsilon \rangle$ & 2.3 GeV$^3$ 
[Eq.(\ref{lepton})] \\
$\langle  O_8^\psi(^3S_1)  \rangle$ &  0.014 GeV$^{3}$
\cite{r:braaten-fleming,r:CGMP,r:cho-leibovich} \\
$\langle  O_8^{\psi'}(^3S_1)  \rangle$ &  0.0042 GeV$^{3}$
\cite{r:braaten-fleming,r:CGMP,r:cho-leibovich} \\
$\langle  O_8^\psi (^1S_0)  \rangle 
\approx \langle  O_8^\psi (^3P_0)  
\rangle /m_c^2$ & $10^{-2}\, {\rm GeV}^3$ \cite{r:cho-leibovich} \\
$\langle O_8^{\chi_{c0}} (^3S_1)  \rangle$ 
& 0.0076 GeV$^3$ \cite{r:CGMP,annrev}\\
$\langle \Upsilon |O_8(^3S_1) |\Upsilon \rangle$
& $5 \times 10^{-4}$ GeV$^3$ [Eq.(\ref{bound})] \\
\hline
Other parameters & Value \\
\hline
$m_c$ & 1.5 GeV \\
$m_b$ & 4.9 GeV\\
$\alpha_s(2m_c)$ & 0.253 \\
$\alpha_s(2m_b)$ & 0.179 \\
\hline
\end{tabular}
\end{table}

\newpage
\section*{Figure Captions}
\begin{enumerate}

\item \label{fig1} 
Some of the contributing Feynman diagrams for the short-distance 
processes:\\
(a) $b \bar b (^3S_1, \underline{1}) \to \gamma^*   \to 
          c \bar c (^{2S+1}L_{J}, \underline{8})  g$;\\
(b) $b \bar b (^3S_1, \underline{1}) \to g^*gg \to   
          c \bar c (^3S_1, \underline{8})  gg$;\\
(c) $b \bar b (^3S_1, \underline{8}) $ $\to  g^*   \to 
              c \bar c (^3P_J, \underline{1}) g$; and \\
(d) $b \bar b (^3S_1, \underline{8}) \to g^*  \to  
	      c \bar c (^3S_1, \underline{1})  gg$.

\item \label{fig2} 
The inclusive energy spectra $d \Gamma /d x_v$ of  $\psi$ 
in the decay  $\Upsilon \to \psi+X$ with the following color-octet 
processes :
$b \bar b (^3S_1, \underline{1}) \to g^*gg \to   
          c \bar c (^3S_1, \underline{8})  gg$ (solid) as 
predicted by Eq.~(\ref{edist1}) compared with the fragmentation approximation 
(dashed), and 
$b \bar b (^3S_1, \underline{8}) \to g^*  \to  
	      c \bar c (^3S_1, \underline{1})  gg$ 
as predicted by Eq.~(\ref{edist2}) multiplied by 300 (dashed-dotted).
\end{enumerate}

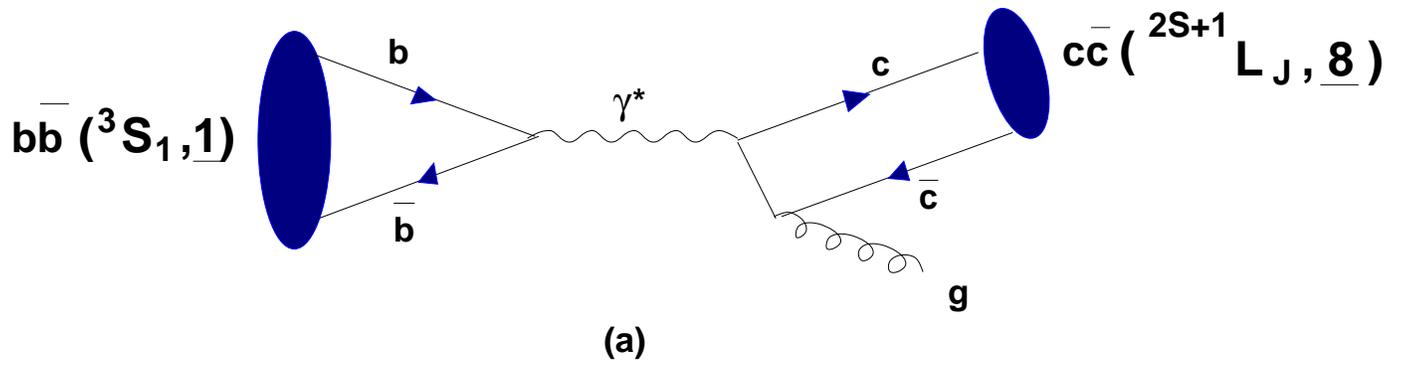
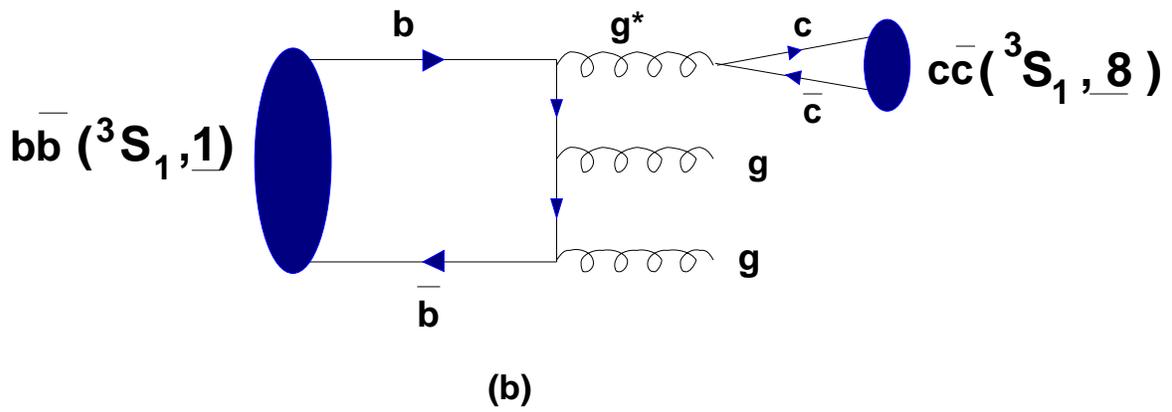
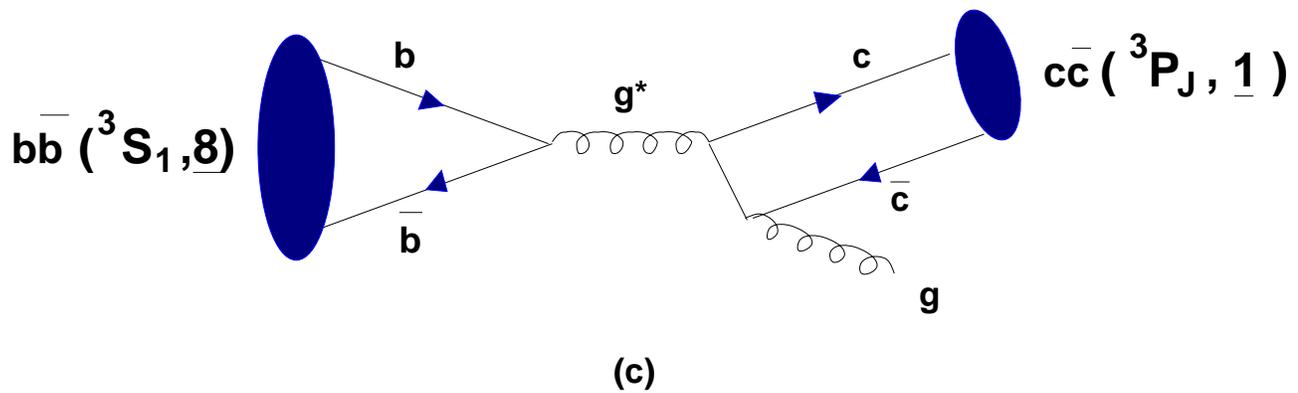
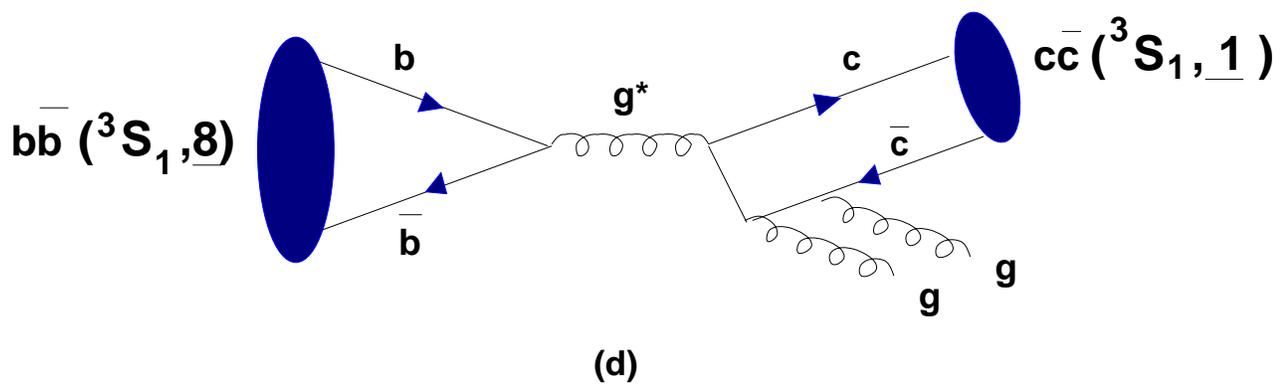
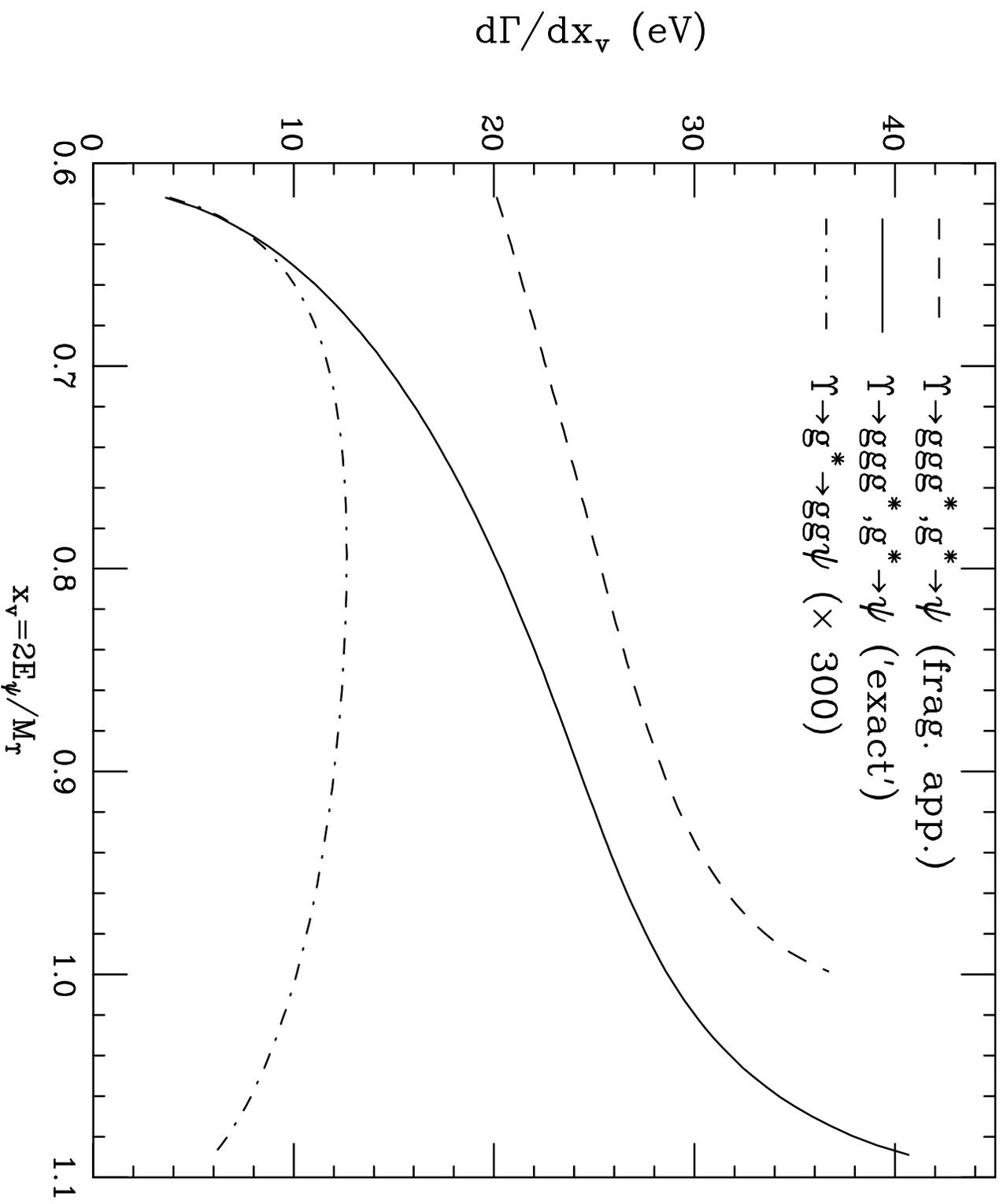
\end{document}